\documentclass{jpp}
\usepackage{graphicx}
\usepackage{xcolor}

\usepackage{hyperref}
\hypersetup{
	breaklinks=true,
	colorlinks=true,
	linkcolor=blue,
	citecolor=purple,
	pdfstartview={FitH},
	pdfview={FitH 0},
	pdfauthor={Jeffrey B. Parker},
 }

\usepackage[utf8]{inputenc}
\usepackage[T1]{fontenc}
\usepackage{lmodern}

\usepackage{amsmath}

\usepackage{bm}

\newcommand{\unitgreek}[1]{\bm{\hat{#1}}}

\providecommand{\eqnref}{}
\renewcommand{\eqnref}[1]{\eqref{#1}}

\newcommand{\ud}{\mathrm{d}}
\newcommand{\ui}{\mathrm{i}}

\renewcommand{\b}{\beta}

\newcommand{\e}{\epsilon}

\newcommand{\g}{\gamma}

\newcommand{\vp}{\varphi}
\renewcommand{\P}{\Phi}

\newcommand{\w}{\omega}

\renewcommand{\d}[2]{\frac{\ud #1}{\ud #2}}						
\renewcommand{\v}[1]{\mathbf{#1}}				
\newcommand{\unit}[1]{{\v{\hat{#1}}}}			
\newcommand{\pd}[2]{\frac{\partial #1}{\partial #2}}		

\newcommand{\vk}{\v{k}}
\newcommand{\vx}{\v{x}}

\usepackage{braket}  

\shorttitle{Topological phase in plasma physics}
\shortauthor{J. B. Parker}

\title{Topological phase in plasma physics}

\author{Jeffrey B. Parker\aff{1}
  \corresp{\email{jeff.parker@wisc.edu}}}
  
\affiliation{\aff{1}Department of Physics, University of Wisconsin--Madison, Madison, WI 53706 USA}

\begin{document}

\maketitle

\begin{abstract}
Recent discoveries have demonstrated that matter can be distinguished on the basis of topological considerations, giving rise to the concept of topological phase.  Introduced originally in condensed matter physics, the physics of topological phase can also be fruitfully applied to plasmas.  Here, the theory of topological phase is introduced, including a discussion of Berry phase, Berry connection, Berry curvature, and Chern number.  One of the clear physical manifestations of topological phase is the bulk-boundary correspondence, the existence of localized unidirectional modes at the interface between topologically distinct phases.  These concepts are illustrated through examples, including the simple magnetized cold plasma.  An outlook is provided for future theoretical developments and possible applications.
\end{abstract}

\section{Introduction}
The aim of this article is to introduce the concepts and physics of topological phase in the context of plasma physics.  The application of topological phase in plasmas is in a fledgling state, although this exciting subject overlaps with active areas of research in other fields of physics.  
 
Broadly speaking, topological phase refers to the notion that a bulk system can be characterized by an integer-valued topological invariant.  More precisely, the topological invariant describes a global property of an eigenfunction in wave vector space.  This type of topology has a more abstract nature than, for instance, the standard topological property of the number of holes of an object in physical space.  An important feature of topological invariants is that they are constrained by topological quantization and are generally not altered under smooth deformations, and so their physical consequences may be robust against perturbations.

A clear physical manifestation of the topological phase arises when two topologically distinct materials are adjacent.  The bulk-boundary correspondence principle states that within a common bandgap at the interface between the two materials, a spatially localized mode exists, referred to as an edge state, or topological wave.  These edge states have attracted interest because of their topological robustness and potential for unidirectional, backscatter-resistant propagation.

The first glimpses of topological phase trace back to the quantization of the Hall conductance in condensed matter systems in the integer quantum Hall Effect \citep{klitzing:1980,laughlin:1981,thouless:1982,avron:1983,simon:1983,niu:1985}.  The conductance of a sample was experimentally measured to occur in integer multiples of $e^2/h$, where $e$ is the elementary charge and $h$ is Planck's constant.  Eventually it was realized that this integer multiple corresponded to a topological invariant called the Chern number that described the sample bulk, with corresponding electron edge states that allowed conduction.

Later it was realized that similar topological phases could be found in photonic crystals  \citep{haldane:2008,raghu:2008}.  This discovery reflects the principle that topological phase is not inherently dependent on quantum mechanics but is a property of waves.  The periodic metamaterial structure of a photonic crystal gives rise to Bloch states and Bloch bands analogous to those in condensed matter systems.    This field of topological photonics may offer novel disorder-robust routes to controlling light \citep{lu:2014,ma:2015,ozawa:2019}.

Topology in condensed matter and photonic systems are studied in systems with an underlying periodic lattice structure.  Some mechanical and acoustic systems in which topological phases and edge states have been explored are also based on periodic lattices \citep{peano:2015,yang:2015,he:2016,nash:2015,huber:2016}.

In contrast, plasmas and fluids are typically described mathematically as a smooth continuum, coarse-grained over the length scale of individual particles.  This distinction gives rise to a very different structure of the wave vector space.  When there is a periodic lattice, the wave vector space is also periodic and can be limited to the first Brillouin zone.  In an infinite continuum medium, the wave vector space extends to infinity.  An important breakthrough was that of \citet{delplace:2017}, who demonstrated that a model in geophysical fluid dynamics can be understood through topological phase and bulk-boundary correspondence.  Other topological phenomena in fluid and continuum electromagnetic media have also been discovered \citep{silveirinha:2015,perrot:2019,souslov:2019,marciani:2020}.

The rich wave physics of plasma makes it likely they can host a variety of topological effects.  Some recent studies have begun to scratch the surface.  For instance, topological properties of a magnetized cold plasma have been studied \citep{gao:2016,parker:2020,fu:2020}.  The Alfv\'{e}n continuum may also be topological in the presence of magnetic shear, leading to a new interpretation of the reversed-shear Alfv\'{e}n eigenmode as a topological edge wave \citep{parker:2020rsae}.  This work also found non-trivial topology in the whistler band within Hall Magnetohydrodynamics.  Yet a systematic study for how topological phase manifests in plasmas and an understanding of the physical consequences and applications are at their inception.

The purpose of this paper is to provide an accessible introduction to these concepts and their applications, without requiring any background in condensed matter physics or differential geometry.  The emphasis is on continuum models with application to plasma physics or geophysical or astrophysical fluids.  For more complete and thorough treatment of topological physics, other reviews may be consulted (e.g., \cite{hasan:2010,ozawa:2019,bernevig:book,asboth:2016}).

We review in section~\ref{sec:background} some essential mathematical background of Berry phase and Chern numbers.  In section~\ref{sec:examples}, we first discuss the shallow-water model for its analytical simplicity, then consider the topological characterization of a magnetized cold plasma and describe a topological wave that may be found at the boundary of a magnetized plasma and vacuum.  In section~\ref{sec:symmetries}, we discuss some important relationships between topology and discrete symmetries.  We provide an outlook in section~\ref{sec:discussion}.

\section{Mathematical background}
\label{sec:background}
In this section we review the mathematical background for topological phase.

\subsection{Berry Phase}
\subsubsection{Discrete Berry Phase}
A Berry phase describes phase evolution of a complex vector as it changes around a closed loop \citep{berry:1984,hannay:1985,berry:1988}.  The Berry phase probes the underlying geometric structure.  A non-zero Berry phase is analogous to the situation of a vector not returning to its original direction when it undergoes parallel transport around a loop on a curved surface. A standard example for where a Berry phase arises is the adiabatic evolution of a quantum mechanical wavefunction.  Berry or geometrical phases have also found numerous applications in plasma physics \citep{littlejohn:1988,liu:2011,liu:2012,brizard:2012,rax:2019,burby:2013}.  To discuss Berry phase in a general way, our setting is a Hilbert space, and we use bra-ket notation, where the Hermitian product of two vectors $\ket{u}$ and $\ket{v}$ is denoted by $\braket{u|v}$.  If $a$ and $b$ are constants, then $\braket{au|bv} = a^* b \braket{u|v}$, and an asterisk denotes complex conjugation.

As is often the case, one can first gain intuition in a discrete setting.  Suppose we have $N$ unit vectors, $\ket{u_1}, \ldots, \ket{u_{N}}$, as depicted in figure~\ref{fig:berry_phase_vectors}(a).  The Berry phase of this sequence of vectors is defined as
	\begin{equation}
		\g = -\Imag \ln \bigl[ \braket{u_1 | u_2} \braket{u_2 | u_3} \cdots \braket{u_{N}| u_1} \bigr].
		\label{discrete_berry_phase1}
	\end{equation}
$\g$ is the Berry phase around a closed loop formed by the discrete sequence.  For a complex number $z = |z|e^{\ui\varphi}$, $\Imag \ln z = \Imag( \ln|z| + \ui \varphi) = \varphi$, so the $\Imag \ln (\cdot)$ operation yields the complex phase and discards the magnitude.  The product of the $N$ inner products of the vectors has some complex phase, and the negative of that phase is the Berry phase.  Since different branch choices of the complex logarithm leads to non-uniqueness of the phase up to integer multiples of $2\pi$, the Berry phase is defined modulo $2\pi$.

Typically, when working with complex unit vectors, their overall phase is arbitrary such that any physical result does not depend on the phase.  The Berry phase is constructed such that it is invariant to these phases.  To see this, consider a gauge transformation induced by phase factors $\beta_j$, where a new set of $N$ vectors is defined
	\begin{equation}
		\ket{u_j} \to e^{-\ui \b_j} \ket{u_j}.
	\end{equation}
The Berry phase computed from the transformed vectors is exactly $\g$ because all of the individual phases cancel out.  The Berry phase is said to be invariant to the gauge transformation, or gauge invariant.  The gauge invariance of Berry phase suggests it may be connected to a physically observable phenomenon.  

\subsubsection{Continuous Formulation of Berry Phase}
\begin{figure}
		\centering
		\includegraphics[scale=0.7]{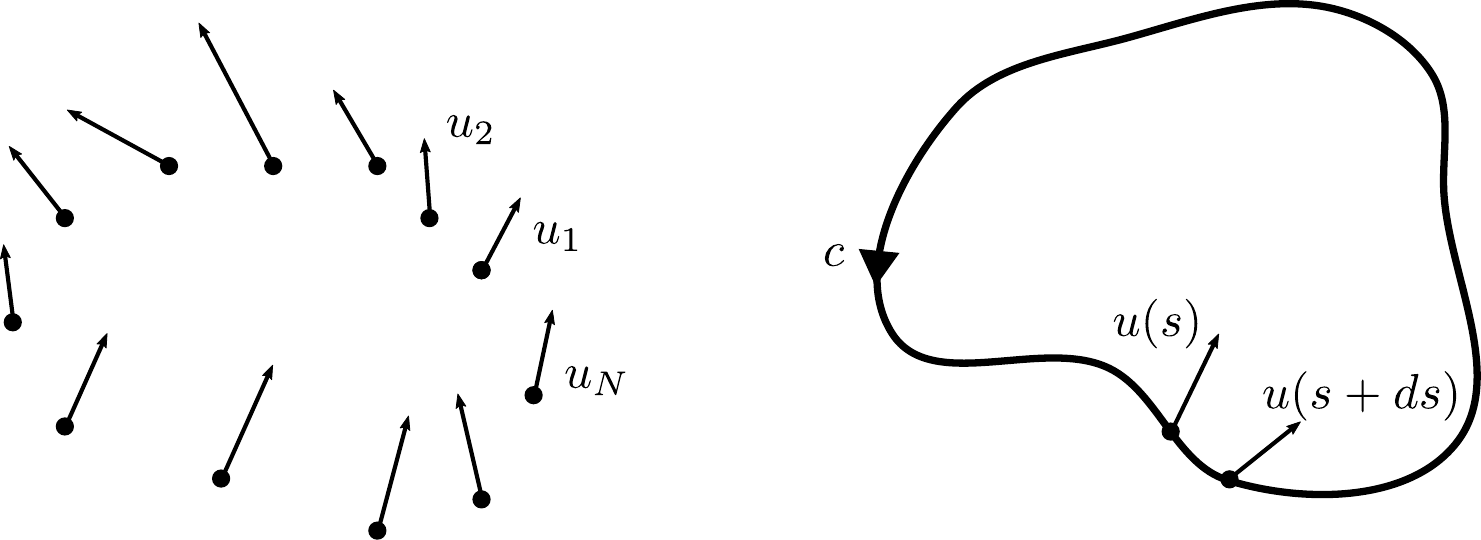}
		\caption{Discrete (left) and continuous (right) vectors for Berry phase, where $c$ is a closed curve.}
		\label{fig:berry_phase_vectors}
	\end{figure}

Let us take the continuum limit of the Berry phase.  We start from the expression
	\begin{equation}
		\g = -\sum_{j=0}^{N-1} \Imag \ln \braket{u_j | u_{j+1}}.
		\label{discrete_berry_phase2}
	\end{equation}
which is equivalent to \eqnref{discrete_berry_phase1} modulo $2\pi$.  We suppose $j$ is an index that parameterizes some property, and we let $j$ pass to the continuous parameter $s$ and $\ket{u_j} \to \ket{u(s)}$ as shown in figure~\ref{fig:berry_phase_vectors}.  An additional constraint is that we impose that $\ket{u(s)}$ be continuous and differentiable.

In the continuum limit, the intuitive notion is to let $\braket{u_j | u_{j+1}} \to \braket{u(s) | u(s + \ud s)}$, but we immediately replace this by setting
	\begin{equation}
		\ket{u(s + \ud s)}= \ket{u(s)} + \ket{u'(s)} \ud s + O(\ud s^2),
	\end{equation}
where $\ket{u'(s)} = \frac{\ud}{\ud s} \ket{u(s)}$ is the tangent vector to $\ket{u(s)}$.  Then,
	\begin{align}
		\braket{u(s) | u(s + \ud s)} &= \braket{u(s) | u(s)} + \braket{u(s) | u'(s)} \ud s + O(\ud s^2) \notag \\
			& = 1 + \braket{u(s) | u'(s)} \ud s + O(\ud s^2).
	\end{align}
Hence,
	\begin{equation}
		\ln \braket{u(s) | u(s + \ud s)} = \braket{u(s) | u'(s)} \ud s + O(\ud s^2).
	\end{equation}
In the continuum limit of \eqnref{discrete_berry_phase2} and taking the sum to an integral, we obtain
	\begin{equation}
		\g = - \Imag \oint \ud s \braket{u(s) | u'(s)},
		\label{continuous_berry_phase1}
	\end{equation}
where the integral is over the closed loop.

An important property is that $\braket{u(s) | u'(s)}$ is pure imaginary, which follows from differentiating $\braket{u(s) | u(s)}=1$ with respect to $s$.  Thus, equivalent to \eqnref{continuous_berry_phase1} is
	\begin{equation}
		\g = \ui \oint \ud s \braket{u(s) | u'(s)}.
		\label{continuous_berry_phase2}
	\end{equation}

\subsection{Berry Connection}
\label{sec:berryconnection}
\begin{figure}
		\centering
		\includegraphics[scale=0.7]{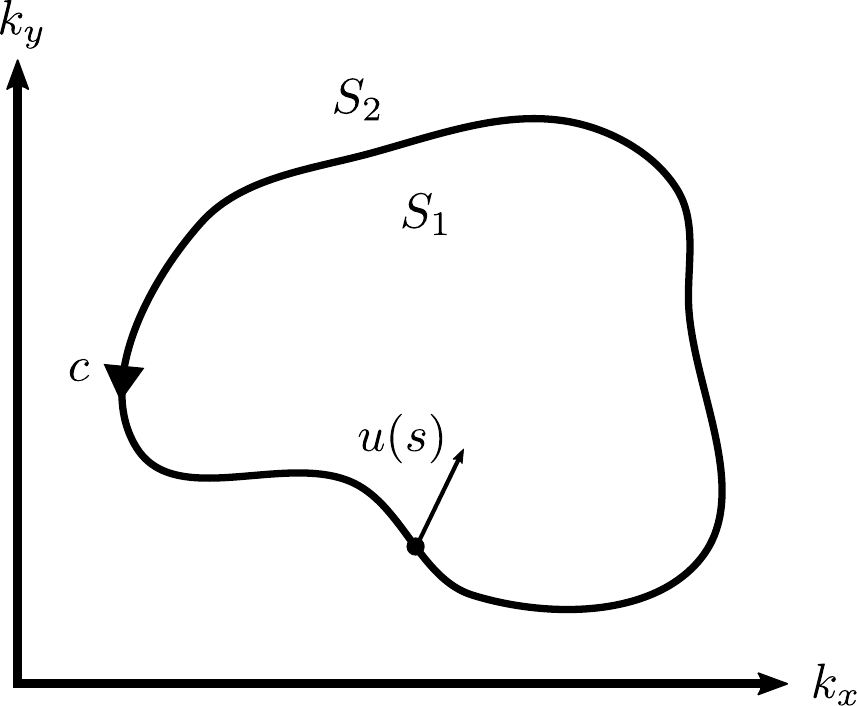}
		\caption{Parameter space $\vk$.  $S_1$ and $S_2$ denote the inside and outside of the closed loop $c$.}
		\label{fig:berry_phase_k_space}
	\end{figure}

Thus far, we have merely dealt with a parameterized loop of unit vectors, with vectors defined only on that loop. Now, suppose there is a two-dimensional (2D) parameter space, with coordinates denoted by $\vk = (k_x, k_y)$ as in figure~\ref{fig:berry_phase_k_space}.  While this parameter space can be completely general, we will primarily be concerned with the situation where the parameters are wave vectors in Fourier space.  Furthermore, we assume one can define vectors $\ket{u(\vk)}$ which exist within some neighbourhood of $c$, not just $c$ itself.  Along the path $c$, we can write $\ket{u(s)} \to \ket{u(\vk(s))}$, and express
	\begin{equation}
		\d{}{s} u(\vk(s)) = \pd{u}{k_j} \d{k_j}{s},
	\end{equation}
where a sum over repeated indices is implied.  The Berry phase can be written
	\begin{equation}
		\g = \ui \oint \ud\vk \cdot \Braket{u(\vk) | \nabla_\vk u(\vk)}.
	\end{equation}
We define
	\begin{equation}
		\v{A}(\vk) = \ui \Braket{u(\vk) | \nabla_\vk u(\vk)},
		\label{def:berry_connection}
	\end{equation}
which is called the Berry connection or Berry potential.  The terminology `connection' comes from differential geometry, whereas the term `potential' arises from an analogy with the vector potential of electromagnetism.  The Berry connection is pure real, as can be seen by taking the gradient of $\Braket{u | u} = 1$ with respect to $\vk$.

Let us consider how the Berry connection and Berry phase transform under a gauge transformation.  Similar to the discrete case, define a gauge transformation to construct a new set of unit vectors differing from the original by a phase
	\begin{equation}
		\ket{u(\vk)} \to e^{-\ui \b(\vk)} \ket{u(\vk)},
	\end{equation}
where the phase $\b(\vk)$ is real and differentiable.  Using \eqnref{def:berry_connection}, the Berry connection transforms as
	\begin{equation}
		\v{A}(\vk) \to \v{A}(\vk) + \nabla_\vk \b(\vk),
		\label{gaugetransform_berryconnection}
	\end{equation}
and therefore the Berry connection is not gauge invariant.  Because the factor $e^{-i \b(\vk)}$ must be single-valued, the Berry phase is gauge invariant modulo $2\pi$.

\subsection{Berry Curvature}
Define the Berry curvature
	\begin{equation}
		\v{F}(\vk) = \nabla_{\vk} \times \v{A}(\vk).
		\label{def:berry_curvature}
	\end{equation}
The curvature encodes information about the local geometric structure.  The curl applies in a 3D (or 2D restriction thereof) setting.  One can formulate the curvature and other concepts here in higher dimensions using differential forms, but we have no need for that machinery at the moment.  For a 2D parameter space $(k_x, k_y)$, there is only one component of the Berry curvature:
	\begin{equation}
		F(\vk) = \ui \left( \Braket{ \pd{u}{k_x} | \pd{u}{k_y}} - \Braket{ \pd{u}{k_y} | \pd{u}{k_x}} \right) = -2\, \Imag \Braket{ \pd{u}{k_x} |  \pd{u}{k_y}}.
		\label{berrycurvature:2parameter}
	\end{equation}

The Berry curvature is gauge invariant.  This fact follows from \eqref{gaugetransform_berryconnection} and that the curl of a gradient vanishes.  Thus there is a suggestive analogy with the magnetic field.  The Berry connection $\v{A}$ is analogous to the vector potential, and is not invariant under a gauge transformation.  The Berry curvature $\v{F}$ is analogous to the magnetic field, and is invariant under a gauge transformation.

\subsection{Chern Theorem}
A simple version of the Chern theorem states that the integral of the Berry curvature over a closed 2D manifold is
	\begin{equation}
		C = \frac{1}{2\pi} \int \ud \v{S} \cdot \v{F}(\vk),
		\label{cherntheorem}
	\end{equation}
for some integer $C$.  Here, $C$ is called the Chern number of the surface.  It is a topological invariant associated with the manifold of states $\ket{u(\vk)}$ defined on the surface.  The Chern number is a property of the collection of complex vectors over the surface, not just the surface itself.  As a topological invariant, the Chern number provides a topological quantization.

That the Chern number must be an integer can be understood intuitively as follows.  Consider again figure~\ref{fig:berry_phase_k_space}.  Let $\v{A}_1$ be the Berry connection constructed with a gauge such that it is smooth on $S_1$, and similarly for $\v{A}_2$ on $S_2$.  The Chern number is given by
	\begin{align}
		C &= \frac{1}{2\pi} \left( \int_{S_1} \ud\v{S} \cdot \v{F}  +  \int_{S_2} \ud \v{S} \cdot \v{F} \right) \\
			&= \frac{1}{2\pi} \left( \oint_c \ud\v{k} \cdot \v{A}_1 - \oint_c \ud\v{k} \cdot \v{A}_2 \right) \\
			&= \frac{1}{2\pi} \left( \g_1 - \g_2 \right).
	\end{align}
Since gauge invariance requires the Berry phases $\g_1$ and $\g_2$ be equal modulo $2\pi$, the Chern number must be an integer.

When the Chern number is non-zero, one cannot construct a smooth, continuous gauge for $\ket{u(\vk)}$ over the entire closed surface.  A related concept is the hairy ball theorem, which states that due to the topology of the sphere, any vector field on the sphere must have singularities or vanishing points.  While there is as yet no direct physical interpretation of the Chern number in continuum systems, the physical interpretation of the Chern number in photonic crystals has been advanced recently; it has been shown the photonic Chern number is related to the thermal fluctuation-induced angular momentum \cite{silveirinha:2019prx,silveirinha:2019}.

The Chern theorem relates geometry and topology.  This is analogous to the Gauss--Bonnet theorem, which relates an integral of a local geometric quantity, the Gaussian curvature, to a global topological quantity, the Euler characteristic.  Here, the Berry curvature is a local geometric quantity, whereas the Chern number is a global topological property.  The analogy is not perfect, however, as an important distinction is that Gaussian curvature reflects a property of the base manifold while the Berry curvature reflects a property of a vector field on the base manifold.

\subsection{Alternative form of the Berry Curvature}
Given some unit vector as a function of two parameters $\vk = (k_x, k_y)$, \eqnref{def:berry_curvature} provides a formula for the local Berry curvature $F(\vk)$.  This standard form depends on derivatives of the vectors at different parameter values, which poses difficulties for numerical computations because great care is required to ensure a smooth gauge.  An alternative form for the Berry curvature which is often useful can be given under the conditions that the parameterized vector arises from a non-degenerate Hermitian eigenvalue problem.  The alternative form is manifestly gauge invariant.

Consider the eigenvalue problem
	\begin{equation}
		H \ket{n} = \w_n \ket{n},
		\label{alternateform:eigenvalueproblem}
	\end{equation}
where $H$ is a Hermitian $N \times N$ matrix, which acts as an effective Hamiltonian.  Thus the $\w_n$ are real.  Here we assume that there is no degeneracy, so that all $N$ eigenvalues are distinct.  A discussion of the degenerate case can be found in \citep{bernevig:book}.  Let $\{ \ket{n} \}$, $n=1, \ldots, N$ be an orthonormal eigenbasis.  The Hamiltonian depends on $\vk$, and therefore so do the eigenvalues and eigenvectors.  From \eqnref{berrycurvature:2parameter}, the Berry curvature corresponding to eigenvector $\ket{n}$ is
	\begin{equation}
		F_n(\vk) = -2\, \Imag \Braket{\pd{n}{k_x} | \pd{n}{k_y}}.
		\label{alternateform:startingBerrycuvature}
	\end{equation}

Apply $\partial / \partial k_i$ to \eqnref{alternateform:eigenvalueproblem}, obtaining
	\begin{equation}
		\pd{H}{k_i} \ket{n} + H \Ket{\pd{n}{k_i}} = \pd{\w_n}{k_i} \ket{n} + \w_n \Ket {\pd{n}{k_i}}.
		\label{alternateform:step2}
	\end{equation}
Act on \eqnref{alternateform:step2} from the left with $\bra{m}$.  For $n \neq m$, and assuming non-degenerate eigenvalues, we obtain
	\begin{equation}
		\Braket{m | \pd{n}{k_i}} = \frac{  \displaystyle \Braket{m | \pd{H}{k_i} | n}} {\w_n - \w_m}.
		\label{alternateform:derivativeprojection}
	\end{equation}
	
Use $I = \sum_m \ket{m} \bra{m}$ in \eqnref{alternateform:startingBerrycuvature} to obtain
	\begin{equation}
		F_n(\vk) = -2\, \Imag \left[ \sum_m \Braket{\pd{n}{k_x} | m} \Braket{m | \pd{n}{k_y}} \right].
	\end{equation}
Consider the $m=n$ term in the sum.  We have previously seen that when $\ket{n}$ is a unit vector, $\braket{n | \pd{n}{k_i}}$ is purely imaginary.  Therefore, $\braket{\pd{n}{k_x} | n} \braket{n | \pd{n}{k_y}}$ is real and does not contribute to the sum.  Using \eqnref{alternateform:derivativeprojection} for the remaining terms, we obtain
	\begin{equation}
		F_n(\vk) = -2\, \Imag  \left[ \sum_{m \neq n} \frac{\displaystyle \Braket{n | \pd{H}{k_x} | m} \Braket{m | \pd{H}{k_y} | n} }{(\omega_n - \omega_m)^2}  \right].
	\end{equation}
This can also be written in the usual form without the explicit imaginary part:
	\begin{equation}
		F_n(\vk) = \ui \sum_{m \neq n} \frac{\displaystyle \Braket{n | \pd{H}{k_x} | m} \Braket{m | \pd{H}{k_y} | n} - \Braket{m | \pd{H}{k_x} | n} \Braket{n | \pd{H}{k_y} | m}}{(\omega_n - \omega_m)^2}.
		\label{alternateform:final_noimagpart}
	\end{equation}
This form of the Berry curvature is manifestly gauge invariant, because any phase on the eigenvectors from a gauge transformation cancels out.  This form can be useful in practice, particularly for numerical computations.  The original form of the Berry curvature is not manifestly gauge invariant.  It contains derivatives of the eigenfunctions.  In contrast, \eqnref{alternateform:final_noimagpart} places the derivative on the Hamiltonian rather than on the eigenfunction and eliminates issues of needing to numerically constrain to a smooth gauge.

\subsection{Bulk-Boundary Correspondence}
One of the most important reasons for the widespread interest in topological phases is the bulk-boundary correspondence, which states that the bulk properties and edge properties of systems are connected.  While our discussion so far has been purely an abstract, mathematical discussion, we now turn to the physical manifestations of topological phase.  As already mentioned, the abstract unit vectors $\ket{u(\vk)}$ discussed previously can represent the eigenfunctions of a Hamiltonian, with dependence on wave vector $\vk$.  By use of a Fourier transform to $\vk$ space, one is implicitly considering an infinite material, or that finite-size systems are sufficiently large.  Chern numbers can be computed for each band in the bulk.

The bulk-boundary correspondence principle states that when two materials with differing topological phases and a common gapped spectrum are brought next to each other, modes localized to the interface and crossing the gap must appear at the interface \citep{hasan:2010}.  The bandgap Chern number for one material is $C_{\text{gap},1} = \sum_{n < n_{\text{gap}}} C_n^{(1)}$, summed over all bands below the bandgap in the first material.  Similarly, for the second material the gap Chern number is $C_{\text{gap},2} = \sum_{n < n_{\text{gap}}} C_n^{(2)}$.  If $C_{\text{gap},1} -  C_{\text{gap},2} \neq 0$, propagating surface modes are present in the gap.  The standard heuristic argument for why modes at the interface must appear is that for a gapped spectrum, the Chern number cannot change across the interface unless the gap closes somewhere at the interface.  Closing the gap is accomplished by the surface mode.  Moreover, the difference in Chern number dictates the number and direction of the propagating surface modes \cite{gangaraj:2017}.

While the conventional understanding just given is often assumed to hold, it is typically proven only for specific model systems \citep{silveirinha:2019prx}.  Additionally, in some cases of continuous-media systems, the bulk-boundary correspondence principle has been found to not apply straightforwardly \citep{gangaraj:2020}.  \citet{tauber:2020} found that the number of edge modes can be boundary-condition dependent, and restoration of the correspondence between Chern numbers and number of edge modes requires a more generalized accounting of possible ghost edge modes.

\subsection{Compactness}
\label{sec:compactness}
The Chern theorem of \eqref{cherntheorem} holds for a closed manifold (a manifold without boundary that is compact).  In condensed matter systems or photonic crystals that have an underlying periodic lattice, the wave vector space of the first Brillouin zone is also periodic, topologically equivalent to a torus, and compact.  The Chern theorem can therefore be applied directly.

However, in continuum models that are typically used in plasmas or fluid dynamics, non-compact wave vector manifolds arise naturally.  The wave vector space extends to infinity; $|\vk| = \infty$ can be thought of as the boundary.  It is therefore important to delve at least a little into the issue of compactness to understand whether and how the Chern theorem is applicable.  To be fully precise here, we distinguish between a Chern number, which is an integer-valued topological invariant, and the integral of the Berry curvature.  For a compact manifold, the Chern theorem guarantees these two values are equal.  For a non-compact manifold, that is not necessarily the case, and in fact one may find non-integer results for the integral of the Berry curvature.  The effects of non-compactness are subtle, and may or may not cause difficulties in any given problem.

For example, one of the frequency bands in the cold plasma model discussed in section~\ref{sec:gpp} has a non-integer integral of the Berry curvature, although the other bands have integer values \citep{parker:2020}.   Moreover, in the shallow-water model discussed in section~\ref{sec:shallowwater}, all frequency bands result in integer-valued Berry curvature integrals.  Interpretation in terms of the bulk-boundary correspondence is unclear when non-integer values are present.

There are various ways of dealing with the lack of compactness in continuum models \citep{silveirinha:2015,tauber:2019,souslov:2019}.  If the problem stems from infinite wave vectors, one method is to introduce a regularization at small scales that enables compactification.  For example, if the behaviour is regularized to decay sufficiently rapidly at large wave vectors, the infinite $\vk$-plane can be mapped onto the Riemann sphere, which is a compact manifold and enables the Chern theorem to apply.  Physically, such a regularization can be justified because the continuum model ceases to be valid at the microscopic scale of the interparticle spacing and the discreteness of the plasma becomes apparent.  Regularization based on plasma discreteness for the cold plasma model was used by \citet{parker:2020}.

Instead of regularizing based on some physically motivated reason, one might try to tackle the lack of compactness directly.  For non-compact manifolds, the index theorems relating an analytical index and topological index can be generalized, and there are additional boundary terms in the index formula \citep{eguchi:1980}.  The boundary data arising from infinite wave vectors can be responsible for the non-integer integral of the Berry curvature.

\section{Examples}
\label{sec:examples}
The concepts described in the previous section are illustrated with specific examples.  The first example, in section~\ref{sec:shallowwater}, comes from the shallow-water equations of geophysical fluid dynamics \citep{delplace:2017}.  This example, although not directly related to plasma physics, is discussed in detail for its analytic transparency, minimal complexity, and clear physical manifestation of the bulk-boundary correspondence principle.  The mathematical framework of wave analysis is the same as commonly used in plasma physics: linearized equations of motion and Fourier analysis.  This example also serves to highlight the interdisciplinary nature of these topological ideas.  In section~\ref{sec:gpp}, we discuss topology of a magnetized cold plasma and describe a topological surface wave between plasma and vacuum.

\subsection{Shallow-water equations and equatorial waves}
\label{sec:shallowwater}
Following \citet{delplace:2017}, the non-dimensionalized, linearized fluid equations of motion of the shallow-water system are
	\begin{subequations}
	\label{shallowwater:system}
	\begin{align}
		\partial_t \eta &= -\partial_x u_x - \partial_y u_y, \\
		\partial_t u_x 		&= -\partial_x \eta  + f u_y, \\
		\partial_t u_y 		&= -\partial_y \eta  - f u_x,
	\end{align}
	\end{subequations}
where $u_x$ and $u_y$ are the fluid velocities and $\eta$ is the perturbation about the mean height.  The $f$-plane model is used here, which is a local model of a rotating sphere using a constant value for the Coriolis parameter $f$ at a particular latitude, and $x$ and $y$ are the coordinates on the tangent plane.  The sign of $f$ changes across the equator from the northern to southern hemisphere.  To facilitate analysis, the $f$-plane is taken to be infinite and homogeneous.  Note that $f$ appears in a manner similar to the cyclotron frequency of charged particles moving in a magnetic field.

\begin{figure}
		\centering
		\includegraphics[scale=0.7]{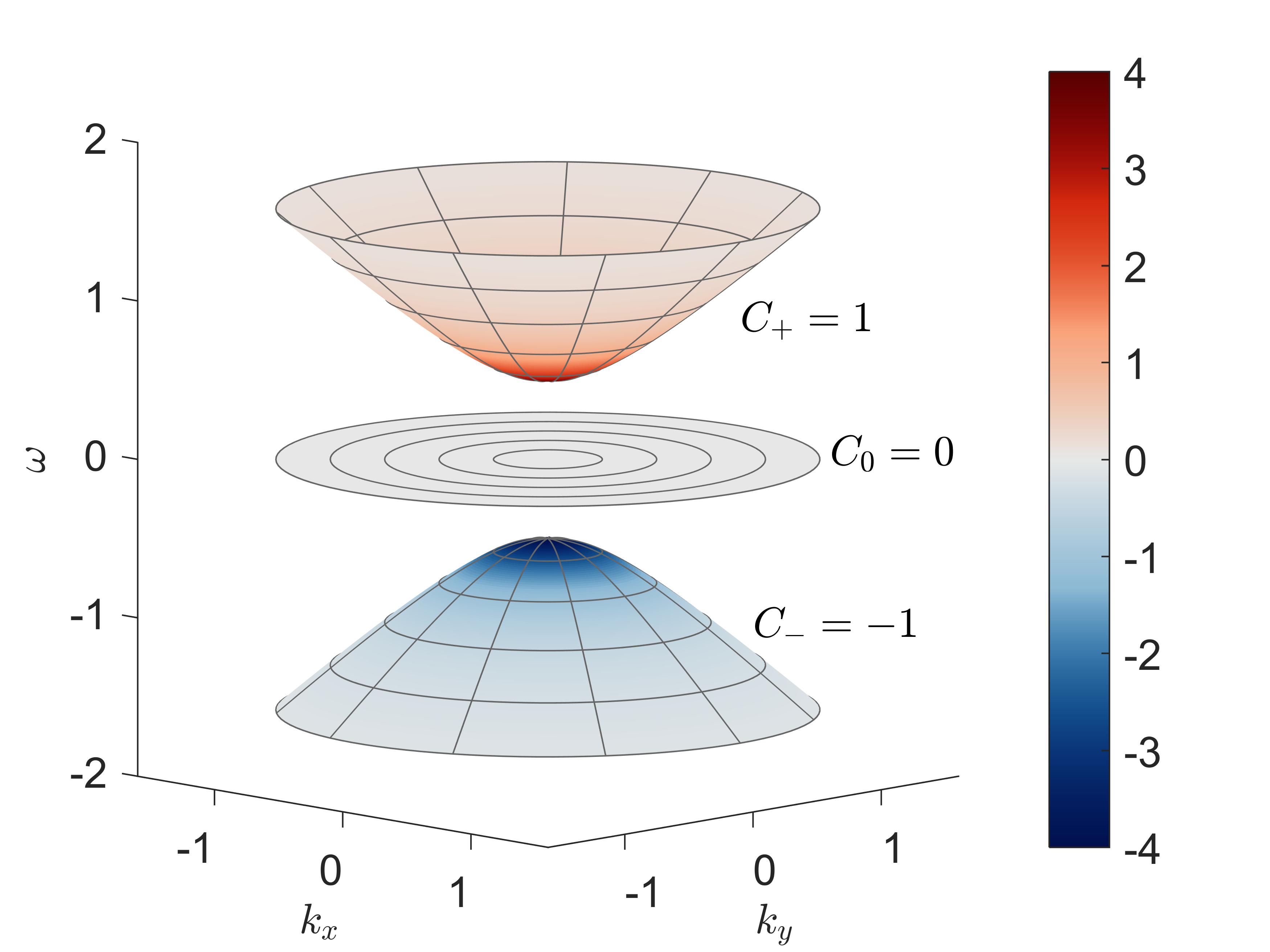}
		\caption{Dispersion relation for the three frequency bands in the shallow-water model.  Colour shows the value of the Berry curvature, and the Chern number of each band is indicated.  Here, $f=0.5$.}
		\label{fig:shallow_water_dispersion}
	\end{figure}

Following standard Fourier analysis, we treat all perturbation quantities as having dependence $e^{\ui (k_x x + k_y y - \w t)}$.  The linearized system can then be written as the eigenvalue equation
	\begin{equation}
		H \ket{\psi} = \w \ket{\psi},
	\end{equation}
where the frequency $\w$ is the eigenvalue,
	\begin{equation}
		\ket{\psi} = \begin{pmatrix} \eta \\ u_x \\ u_y \end{pmatrix},
	\end{equation}
and
	\begin{equation}
		H = \begin{bmatrix}
		0 & k_x & k_y \\
		k_x & 0 & \ui f \\
		k_y & -\ui f & 0
		\end{bmatrix}.
	\end{equation}
The effective Hamiltonian $H$ is Hermitian.  The eigenvalues are $\w_{\pm} = \pm \sqrt{k^2 + f^2}$ and $\w_0 = 0$, where $k^2 = k_x^2 + k_y^2$.  These modes are the Poincar\'{e} waves and a degenerate zero-frequency Rossby wave.  The non-normalized eigenfunctions are
	\begin{equation}
		\ket{\psi_\pm} = \begin{pmatrix} k^2 \\ \pm k_x \sqrt{k^2 + f^2} + \ui f k_y  \\ \pm k_y \sqrt{k^2 + f^2} -\ui f k_x  \end{pmatrix}, \qquad \ket{\psi_0} = \begin{pmatrix}
			f \\ -\ui k_y \\ \ui k_x
		\end{pmatrix}.
	\end{equation}
The three frequency bands are shown in figure~\ref{fig:shallow_water_dispersion}.

Using the concepts developed in section~\ref{sec:background}, we are now in a position to calculate the Berry connection, Berry curvature, and Chern number of each band.  We show the computation in detail for the $\omega_+$ band; the other two bands are analogous.  The standard inner product is used.

First we compute the Berry connection for this band.  An equivalent expression to \eqref{def:berry_connection} for a non-normalized eigenfunction is
	\begin{equation}
		\v{A}_+(\vk) = -\frac{\Imag \Braket{\psi_+ | \nabla_\vk \psi_+}}{\Braket{\psi_+ | \psi_+}}.
	\end{equation}
It is convenient to express $\nabla_\vk$ in polar coordinates, where $k_x = k \cos \vp$ and $k_y = k \sin \vp$.  Thus,
	\begin{equation}
		\ket{\psi_+} = \begin{pmatrix}
			k^2 \\
			k\sqrt{k^2 + f^2} \cos \vp  + \ui f k \sin \vp \\
			k \sqrt{k^2 + f^2} \sin \vp -\ui f k \cos \vp
		\end{pmatrix},
	\end{equation}
and
	\begin{equation}
		\nabla_\vk \ket{\psi_+} = \unit{k} \begin{pmatrix}
			2k \\
			\frac{2k^2 + f^2}{\sqrt{k^2 + f^2}} \cos\vp  + \ui f \sin \vp \\
			\frac{2k^2 + f^2}{\sqrt{k^2 + f^2}} \sin\vp  -  \ui f \cos \vp
		\end{pmatrix}  +  
		\frac{\unitgreek{\vp}}{k} \begin{pmatrix}
			0 \\
			-k \sqrt{k^2 + f^2} \sin \vp + \ui f k \cos \vp \\
			k \sqrt{k^2 + f^2} \cos \vp  +  \ui f k \sin \vp
		\end{pmatrix}.
	\end{equation}
The $\unit{k}$ component of $\Braket{\psi_+ | \nabla_\vk \psi_+}$ is real and hence does not contribute to $\v{A}_+$.  We obtain
	\begin{equation}
		\v{A}_+(\vk) = -\frac{f}{k \sqrt{k^2 + f^2}} \unitgreek{\vp}.
	\end{equation}

The Berry curvature is
	\begin{equation}
		F_{+}(\vk) = \frac{f}{(k^2 + f^2)^{3/2}}.
	\end{equation}
The Chern number of this band is
	\begin{equation}
		C_+ = \frac{1}{2\pi} \int \ud^2 k\, F_{+} = \operatorname{sign}(f).
	\end{equation}
Topological quantization has emerged: the Chern number can only take on integer values.  The breaking of time-reversal symmetry by the rotation of the Earth results in a topologically non-trivial bulk fluid in the $f$-plane model of the shallow-water system.  One can similarly show that the Chern numbers for the other bands are $C_0 = 0$ and $C_- = -\operatorname{sign}(f)$.  The Chern numbers are indicated in figure~\ref{fig:shallow_water_dispersion}.

This example offers a clear demonstration of the bulk-boundary correspondence principle.  The direct implication is that if $f$ in \eqref{shallowwater:system} is a function of $y$ rather than constant, then there must be a unidirectional wave localized to the spatial region around $f=0$ that spans the frequency gap.  Because the change in Chern number across such an interface is $C_+(f > 0) - C_+(f < 0) = 2$, there are two localized waves.  

The more physical case of interest is an actual spherical surface without the Cartesian approximation.  The Coriolis parameter $f$ changes sign across the equator.  Hence, the equator forms an interface dividing the topologically distinct northern and southern hemispheres.  In fact, the two expected waves guaranteed by the bulk-boundary correspondence principle are the well-known equatorially trapped modes, the Kelvin wave and the Yanai wave \citep{delplace:2017}.  The dispersion relation for both the Kelvin and Yanai wave is monotonic, indicating group velocities of unidirectional, eastward travelling waves.  

Despite the fact that the Cartesian $f$-plane neglects spherical curvature, which is an order-unity effect, analysis of the $f$-plane has yielded the key topological insight that the northern and southern hemisphere are topologically distinct.  Kelvin waves have been clearly observed in the spectrum of fluctuations in the Earth's atmosphere \citep{wheeler:1999}.  In simulations, \citet{delplace:2017} found that equatorially trapped Kelvin waves lying in the frequency gap experienced reduced scattering against static perturbations compared with modes not in the frequency gap, a signature of topological protection.

Although the infinite $\vk$-plane is not compact, the behaviour at infinite $\vk$ has not in this case spoiled the result of finding an integer Chern number by integrating the Berry curvature.  The compactness issue was handled in an alternate way by \citet{delplace:2017}, who considered a 2D compact surface, a sphere, within the 3D parameter space $(k_x, k_y, f)$.  The Berry curvature within the 3D parameter space is that of a monopole at the origin, and hence any closed surface containing the origin will yield the same Chern number.  This calculation can be reconciled with the one presented above by considering a cylinder centred at the origin of finite height in $f$ and very large radius in the $(k_x, k_y)$-plane.  The Berry flux through the side of the cylinder vanishes, and the flux through one end of the cylinder is equal to the flux through the infinite $\vk$-plane at constant $f$ above.  Yet another way of dealing with compactness is through the addition of odd viscosity \citep{souslov:2019,tauber:2019}.

\subsection{Magnetized cold plasma and the gaseous plasmon polariton}
\label{sec:gpp}
In this section, we examine a simple magnetized cold plasma and show that it can host topological phases along with related interface modes.  The magnetic field breaks time-reversal symmetry.

Consider an infinite, homogeneous, ion-electron plasma.  When considering high-frequency electromagnetic waves, it is appropriate to treat the ions as a fixed neutralizing background and only consider electron motion.  The mathematical description of a cold plasma consists the electron equation of motion and Maxwell's equations:
	\begin{subequations}
	\label{homogeneous_cold_plasma_equations}
	\begin{align}
		\pd{\v{v}}{t} &= -\frac{e}{m_e} (\v{E} + \v{v} \times \v{B}_0), \\
		\pd{\v{E}}{t} &= c^2 \nabla \times \v{B} + \frac{e n_e}{\e_0} \v{v}, \\
		\pd{\v{B}}{t} &= -\nabla \times \v{E},
	\end{align}
	\end{subequations}
where $\v{v}$ is the electron velocity, $\v{E}$ the electric field, $\v{B}_0 = B_0 \unit{z}$ the background magnetic field, $\v{B}$ the perturbation magnetic field, $n_e$ the background electron density, $m_e$ the electron mass, $c$ the speed of light, and $\epsilon_0$ the permittivity of free space.

We consider a fixed $k_z$ parallel to the background magnetic field and choose a two-dimensional parameter space $(k_x, k_y)$.  After proper non-dimensionalization and Fourier analysis, one obtains the Hermitian eigenvalue problem $\w \ket{\psi} = H \ket{\psi}$, where $H$ is a $9\times 9$ matrix and $\ket{\psi} = [\v{v}, \v{E}, \v{B}]$.

	\begin{figure}
		\centering
		\includegraphics{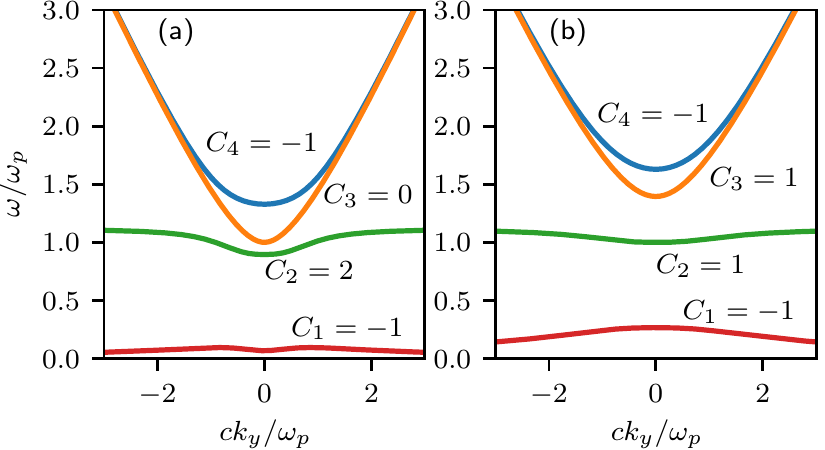}
		\caption{Spectrum of a magnetized, homogeneous cold plasma as a function of $k_y$  ($k_x$ set to zero, but the system is isotropic in the $xy$-plane), where only electron motion is retained.  The two cases show (a) $k_z < k_z^*$ and (b) $k_z > k_z^*$, where $k_z^*$ is a critical point at which a topological transition occurs.  The Chern numbers of the positive-frequency bands are shown \citep{parker:2020}.}
		\label{fig:homog_coldplasma_spectrum}
	\end{figure}

Figure~\ref{fig:homog_coldplasma_spectrum} shows the four positive-frequency bands.  Non-trivial topology is found in multiple bands, as indicated by the non-zero Chern numbers \citep{parker:2020,fu:2020}.  Unlike the shallow-water example, here the straightforward integration of the Berry curvature yields a non-integer result for one of the bands.  As discussed in section~\ref{sec:compactness}, this stems from a lack of compactness.  To obtain the integer Chern numbers shown in figure \ref{fig:homog_coldplasma_spectrum}, a large-wavenumber cutoff of the plasma response was introduced to regularize the small-scale behaviour, motivated by the physical fact that the continuum description breaks down at the scales of the interparticle spacing.

When the topologically non-trivial plasma is placed next to the trivial vacuum, bulk-boundary correspondence suggests the existence of modes at the interface.  One can consider semi-infinite planar system, where the plasma and vacuum each occupies half of the space \citep{yang:2016}.  A more physically realizable system is a confined cylindrical plasma with a radially decreasing density, transitioning to a low-density vacuum-like region.  \citet{parker:2020} investigated this system and demonstrated the existence of topological boundary waves.  An important component of that study was accounting for a finite width of the density interface.  A gaseous plasma cannot sustain a discontinuous density interface with vacuum, and the interface width is typically limited by classical or turbulent diffusion processes.  A discontinuous step in density serves as a good first approximation but is quantitatively limited.  Because the width of the density interface may be comparable in size to the wavelength of the wave, a quantitative treatment is necessary to accurately determine whether the wave can exist.

\begin{figure}
		\centering
		\includegraphics{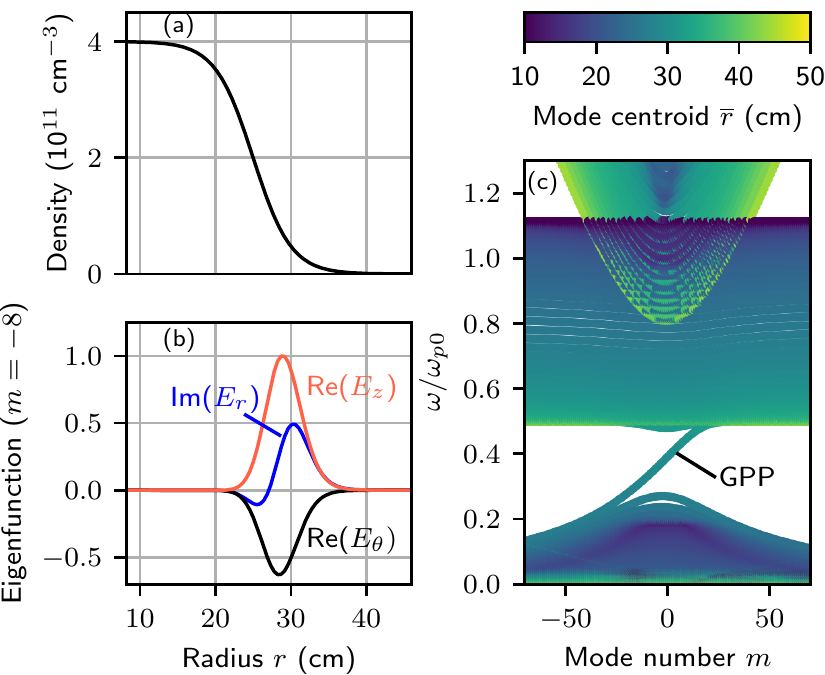}
		\caption{Spectrum of a cylindrical, inhomogeneous, magnetized plasma.  The eigenvalue differential equation in radius was solved using the plasma density as a function of radius shown in (a).  (b) Non-zero components of GPP electric field at azimuthal mode number $m=-8$.  (c) Spectrum as a function of $m$.  The GPP dispersion relation is indicated and crosses the frequency bandgap \citep{parker:2020}.}
		\label{fig:inhomog_coldplasma_spectrum}
	\end{figure}

Figure~\ref{fig:inhomog_coldplasma_spectrum} shows the surface mode at the plasma-vacuum interface.  This mode is the gaseous plasmon polariton (GPP), named for its similarity to surface plasmon polaritons occurring at the surface of metals.  The spectrum of the inhomogeneous plasma was computed by solving the differential eigenvalue equation in radius.  In the figure, the dispersion relation of the GPP is unidirectional and crosses the bandgap.  In \citep{parker:2020}, a typo led to the GPP being described as ``undirectional'' rather than the correct ``unidirectional.''  The GPP can exist in planar as well as cylindrical geometries.

This study also showed that the GPP can be realized in plasma regimes achievable in laboratory experiments.  The parameters used by \citet{parker:2020} were directly motivated by the plasma parameters of the Large Plasma Device \citep{gekelman:2016}.  In this case, a peak plasma density of $n=4 \times 10^{11}$ m$^{-3}$, magnetic field $B=0.1$ T, and density scale length of $L_n \approx 5$ cm were used, and the GPP was calculated to have a frequency of $\sim 2$ GHz.  Hence, the GPP offers a window into the experimental study of topological phenomena in plasma systems.  

\section{Discrete symmetries and topology}
\label{sec:symmetries}
There is a profound relation between symmetry and topology.  In this context, discrete symmetries such as parity and time play a crucial role.  In recent literature, $\mathcal{PT}$ symmetry analysis has been discussed \citep{bender:2007,qin:2019}.  Here, we discuss the consequences of discrete symmetries on the topology of the system.

Parity $\mathcal{P}$ refers to the inversion of one or more spatial dimensions, and time reversal is denoted by $\mathcal{T}$.  Mathematically, when acting on the state vector of a linear system, $\mathcal{P}$ is a linear, unitary operator with $\mathcal{P}^2 = 1$.  The action of parity is here defined as reversing the sign of $\v{x}$ and then applying a unitary operator $V$.

In quantum mechanics, time reversal requires complex conjugation of the wavefunction.  Complex conjugation does not inherently enter in the context of classical physics described in terms of real variables.  However, in the Fourier domain, one can draw a connection to complex conjugation by using the reality condition, which is related to particle-hole conjugation in quantum contexts.  From this perspective, for a linear system, time reversal $\mathcal{T}$ is an antilinear operator with $\mathcal{T}^2 = 1$.  Antilinear means $\mathcal{T} c\ket{\psi} = c^* \mathcal{T} \ket{\psi}$ where $c$ is a complex constant.  The action of time reversal is here defined as mapping $t \mapsto -t$, applying complex conjugation by using the reality condition, and then applying a unitary operator $U$.

Let $H\bigl(\hat{k}\bigr)$ be the matrix differential operator corresponding to the linear problem, which depends on the spatial derivatives, written in terms of $\hat{k}_j = -\ui \partial / \partial x_j$ .  For an infinite, homogeneous system, an eigenmode with wave vector $\vk$ and eigenfrequency $\w$ has the form
	\begin{equation}
		\ket{\phi} = e^{\ui(\vk \cdot \vx - \w t)} \ket{\psi},
	\end{equation}
where $\ket{\psi}$ is just a constant vector that has no space or time dependence, such that
	\begin{equation}
		\w \ket{\phi} = H\bigl( \hat{k} \bigr) \ket{\phi} = H\bigl( \hat{k} \bigr)  e^{\ui(\vk \cdot \vx - \w t)} \ket{\psi} = H\bigl( \vk \bigr)  e^{\ui(\vk \cdot \vx - \w t)} \ket{\psi}.
	\end{equation}
In this expression, $H(\vk)$ is simply $H\bigl(\hat{k}\bigr)$ where $\hat{k}$ has been replaced by the algebraic quantity $\vk$, and $H(\vk)$ is a matrix with no differential operators.  Hence, this leads to the conventional Fourier-space eigenvalue formulation,
	\begin{equation}
		H(\vk) \ket{\psi} = \w \ket{\psi}.
	\end{equation}

A consequence of a parity symmetric system $[H, \mathcal{P}]=0$ is that given one eigenmode $\ket{\phi}$, $\mathcal{P} \ket{\phi}$ is also an eigenmode with the same frequency $\w$.  The proof is simple:
	\begin{equation}
		H\bigl( \hat{k} \bigr) \mathcal{P} \ket{\phi} = \mathcal{P} H\bigl( \hat{k} \bigr) \ket{\phi} = \mathcal{P} \w \ket{\phi} = \w \mathcal{P} \ket{\phi}.
	\label{Psymmetry_eigenmode1}
	\end{equation}
Furthermore,
	\begin{equation}
		\mathcal{P} \ket{\phi} = \mathcal{P} e^{\ui(\vk \cdot \vx - \w t)} \ket{\psi} = e^{\ui(\vk \cdot \mathcal{P} \vx - \w t)} V \ket{\psi} = e^{\ui(\mathcal{P} \vk \cdot \vx - \w t)} V \ket{\psi}.
		\label{Psymmetry_eigenmode2}
	\end{equation}
In the last equality, we have used that if $\mathcal{P}$ flips the sign of one or more components of $\vx$, then one can equally well think of it as flipping the sign of the corresponding components of $\vk$.  Therefore, $\mathcal{P} \ket{\phi}$ is indeed an eigenmode; it has wave vector $\mathcal{P}\vk$, frequency $\w$, and components $V \ket{\psi}$.

We can also consider the consequence of the system being symmetric under time reversal.  A consequence of $[H, \mathcal{T}]=0$ is that $\mathcal{T} \ket{\phi}$ is an eigenmode with frequency $\w^*$.  To see this, note
	\begin{equation}
		H\bigl( \hat{k} \bigr) \mathcal{T} \ket{\phi} = \mathcal{T} H\bigl( \hat{k} \bigr) \ket{\phi} = \mathcal{T} \w \ket{\phi} = \w^* \mathcal{T} \ket{\phi}.
		\label{Tsymmetry_eigenmode1}
	\end{equation}
Using our convention for the action of $\mathcal{T}$, we find
	\begin{equation}
		\mathcal{T} \ket{\phi} = \mathcal{T} e^{\ui(\vk \cdot \vx - \w t)} \ket{\psi} = e^{\ui(-\vk \cdot \vx - \w^* t)} U\ket{\psi^*},
		\label{Tsymmetry_eigenmode2}
	\end{equation}
where the notation $\ket{\psi^*}$ means the complex conjugate of the components is taken.  Hence, $\mathcal{T}\ket{\phi}$ has wave vector $-\vk$, frequency $\w^*$, and components $U\ket{\psi^*}$.

Parity symmetry has direct implications for the Berry connection.  We assume a non-degenerate situation where $\ket{\phi}$ and $\mathcal{P} \ket{\phi}$ can be labelled as part of the same eigenmdode branch.  Here, suppose $\mathcal{P}$ represents full inversion symmetry with $\mathcal{P} \vk = -\vk$.  From \eqref{Psymmetry_eigenmode2}, we have
	\begin{equation}
		\ket{\psi(-\vk)} = V \ket{\psi(\vk)},
		\label{Psymmetry_negativemode}
	\end{equation}
To be more precise, one can slightly generalize \eqref{Psymmetry_negativemode} with a possibly $\vk$-dependent phase factor, which amounts to a gauge transformation as discussed in section \ref{sec:berryconnection}.  From $\v{A}(\vk) = \ui \Braket{\psi(\vk) | \nabla_\vk \psi(\vk)}$, we observe that
	\begin{align}
		\v{A}(-\vk) &= \ui  \Braket{\psi(-\vk) | \nabla_\vk \psi(-\vk)} \label{A_of_minus_k} \\
							&= -\ui \bra{\psi(\vk)} V^\dagger V \ket{\nabla_\vk \psi(\vk)} \\
							&= -\v{A}(\vk),
	\end{align}
up to a gauge transformation, where $\dagger$ denotes the Hermitian adjoint.  Full inversion symmetry then implies the Berry curvature $\v{F}(\vk) = \nabla_\vk \times \v{A}$ is even,
	\begin{equation}
		\v{F}(-\vk) = \v{F}(\vk).
	\end{equation}
	
Time-reversal symmetry can be analysed following a similar approach.  From \eqref{Tsymmetry_eigenmode2}, we have $\ket{\psi(-\vk)} = U\ket{\psi(\vk)^*}$ up to a gauge transformation.  Using this result in \eqnref{A_of_minus_k}, we obtain
	\begin{align}
		\v{A}(-\vk) &= -\ui \bra{\psi(\vk)^*} U^\dagger U \ket{\nabla_\vk \psi(\vk)^*} \\
							&= -\ui \Braket{\psi(\vk)^* | \nabla_\vk \psi(\vk)^* }.
	\end{align}
Because the Berry connection is real, we may take the complex conjugate without changing the result, leading to
	\begin{equation}
		\v{A}(-\vk) = \ui \Braket{\psi(\vk) | \nabla_\vk \psi(\vk) }.
	\end{equation}
One concludes that under time-reversal symmetry, the Berry connection is even and the Berry curvature is odd,
	\begin{gather}
		\v{A}(-\vk) = \v{A}(\vk), \\
		\v{F}(-\vk) = -\v{F}(\vk).
	\end{gather}
Recalling the Chern number $C=(2\pi)^{-1} \int \ud \vk\, F(\vk)$, we see that invariance under $\mathcal{T}$ implies a vanishing Chern number.  Moreover, invariance under both full inversion and time-reversal symmetry implies the Berry curvature itself vanishes.

\section{Discussion}
\label{sec:discussion}
We have introduced topological band theory in the context of plasmas.  One clear physical manifestation of non-trivial topological phase is the presence of modes occurring at the interface between topologically distinct materials, such as a magnetized plasma and vacuum.  Topological physics, along with many generalizations and extensions not presented here, have been systematically studied and applied in condensed matter and photonics as well as other fields of physics.  For instance, topological classifications beyond the Chern number exist, such as the $\mathbb{Z}_2$ invariant of topological insulators \citep{kane1:2005,kane2:2005}.

Various important effects in plasmas can take one beyond the simple topological band theory discussed in this article.  Topological physics is most well understood in the case of Hermitian Hamiltonians.  In contrast, a non-Hermitian Hamiltonian can occur, for instance, when a system experiences gain and/or loss.  In plasmas or fluids, non-Hermiticity might also arise from the presence of flow shear or a density gradient.  In recent years there has been a significant effort to generalize topological band theory to non-Hermitian Hamiltonians \citep{esaki:2011,hu:2011,leykam:2017,shen:2018,gong:2018,kunst:2018,martinezalvarez:2018}.  A natural jumping off point is the space of $\mathcal{PT}$-symmetric Hamiltonians, which can under certain circumstances give real spectra like a Hermitian Hamiltonian.  In the general case with complex eigenvalues and exceptional points, topological classification, topological protection, and the relation to bulk-boundary correspondence are still not fully settled, although much progress has been made in specific problems.  Understanding non-Hermitian topological effects in plasmas and fluids is an open area.

Nonlinearity is another important feature of plasmas and fluid systems.  The theory in terms of an effective Hamiltonian and frequency bands is based on a linearization around equilibrium, an assumption that may have limited validity in many situations.  Effects from nonlinearity have been studied in topological photonics \citep{lumer:2013,leykam:2016,smirnova:2020}.  In plasmas, the interplay between topology and nonlinearity is ripe for exploration.  

Further investigations will deepen our understanding of the physics of topological phase in plasmas and uncover the behaviour of topological modes in plasmas.  Significant theoretical development is needed to unravel the topological nature of the diversity of plasmas at different parameter regimes and scales.  Laboratory investigations are within reach to probe experimental consequences and uses of topological physics in plasmas.  Potential applications of this emerging area include the ability to predict new interface modes using the bulk-boundary correspondence.  The presence of these modes in some circumstances might be used to provide diagnostic information on plasma parameters, or in other situations might provide new means of exerting control over plasmas.  Topological plasma waves may also be robust to perturbations.

\begin{acknowledgments}
\textbf{Acknowledgments.} Useful discussions with Brad Marston and Pierre Delplace are acknowledged.
\end{acknowledgments}

\textbf{Funding.} Support for this research was provided by the University of Wisconsin-Madison, Office of the Vice Chancellor for Research and Graduate Education with funding from the Wisconsin Alumni Research Foundation.

\textbf{Declaration of Interests.}  The author reports no conflict of interest.

\textbf{Author ORCID.} J.\ B.\ Parker, orcid.org/0000-0002-9079-9930


\end{document}